\documentclass[onecolumn,showpacs,preprintnumbers,amsmath,amssymb,prl,preprint]{revtex4}

\usepackage{graphicx}
\usepackage{dcolumn}
\usepackage{bm}

\begin{document}
\title{Backaction Dephasing by a Quantum Dot Detector}

\author{Toshihiro Kubo$^{1,2}$}
 \email{kubo.toshihiro.ft@u.tsukuba.ac.jp}
\author{Yasuhiro Tokura$^{1,2}$}%
 \affiliation{%
$^1$Graduate School of Pure and Applied Sciences, University of Tsukuba, Tsukuba, Ibaraki 305-8571, Japan\\
$^2$NTT Basic Research Laboratories, NTT Corporation, Atsugi-shi, Kanagawa 243-0198, Japan}%

\date{\today}

\begin{abstract}
We derive an analytical expression for the backaction dephasing rate, which characterizes the disturbance induced by coupling with an environment containing a quantum dot detector (QDD). In this letter, we show that charge noise induces backaction dephasing in an explicit form. In the linear transport regime through a QDD, this backaction dephasing induced by charge noise can be explained as a relaxation by an inelastic electron-electron scattering in Fermi liquid theory. In the low bias voltage regime, the increase or decrease of dephasing rate depends on the QDD energy level, the linewidth functions, and how to apply the bias voltage. Unlike quantum point contact, the dephasing rate would be insensitive to the bias voltage in a high bias voltage regime because of the saturation of charge noise in a QDD.
\end{abstract}

\pacs{03.65.Yz, 73.23.-b, 73.63.Kv}

\maketitle
The Heisenberg uncertainty principle states that a measurement necessarily changes the quantum state \cite{heisenberg,qc}; such a measurement induced disturbance is termed \textit{quantum backaction}. It is important to understand the dephasing process in the context of quantum information processing since the measurement of quantum states is a necessary element of quantum feedback and quantum computing. Understanding the backaction mechanism is significant to control the quantum state coherently.

Quantum dots (QDs) have attracted the attention of many physicists as possible building blocks for application to quantum information processing \cite{loss}. In particular, we focus on the QD as a detector that allows us to readout the charge states of another QD. Backaction dephasing in a charge readout of a QD caused by a quantum point contact (QPC) has been extensively examined both theoretically \cite{which-path1,which-path2,which-path3,which-path4,which-path5} and experimentally \cite{which-path6,which-path7,which-path8}. It is well-known that the current shot noise of QPC detectors induces backaction dephasing. A QD detector (QDD) is used as another type of charge detector \cite{qdd1,qdd2,qdd3,qdd4,qdd5} since it is expected to provide high operating speeds and very high sensitivity. The backaction in a qubit readout by a QDD was studied \cite{set1,set2}. In Ref. \onlinecite{set1}, the backaction for a Cooper-pair  confined in a single Cooper-pair box was investigated, and it was shown that the relaxation rate of a qubit is proportional to the voltage fluctuation at the frequency corresponding to the qubit transition. Understanding the backaction mechanism of a QDD is also useful since a qubit made of QDs is capacitively affected by neighboring QDs constructing a qubit system. For the measured system with the QD that couples to the reservoir, QDD induced backaction dephasing has been investigated theoretically \cite{dephasing-qdd1,dephasing-qdd2,kubo-dephasing}. Although it has been pointed out that in contrast to QPC detectors the backaction dephasing is not associated with the current shot noise and is induced by the charge noise of a QDD \cite{dephasing-qdd1}, the physical origin of the backaction dephasing induced by a QDD was not clarified in explicit form. Moreover, discussions in Ref. \onlinecite{dephasing-qdd1} is applicable only in the limited bias voltage condition.

In this letter, we study the backaction dephasing induced by coupling with an environment containing a QDD to address the following two main issues: (i) What is the physical origin of the backaction dephasing by a QDD? To clarify this question, we derive the analytical expression for the backaction dephasing rate within the framework of nonequilibrium second-order many-body perturbation theory. We show that charge noise of a QDD causes the backaction dephasing. In Refs. \onlinecite{set1} and \onlinecite{set2}, the backaction dephasing in a charge qubit readout by a QDD was already investigated for a Cooper-pair confined in a single Cooper-pair box, namely the measured system has a discrete level. In contrast to this, we discuss the backaction dephasing for delocalized electron in a measured system that has a coupling with the reservoirs. (ii) The other main issue is the difference between the bias voltage dependences of backaction dephasing rate for a QDD and QPC. To discuss this point, we examine the first-order expansion coefficient for backaction dephasing rate expressed as a polynomial function of a bias voltage. Unlike QPC, in the low bias voltage regime, the backaction dephasing rate may decrease by the condition for the QDD energy level, the linewidth functions, and how to apply the bias voltage. Moreover, we show that the backaction dephasing rate is saturated for the high bias voltage. Such behavior can be verified in terms of the visibility of the Aharonov-Bohm (AB) oscillations by changing the bias voltage across the QDD.

\begin{figure}
\includegraphics[scale=0.33]{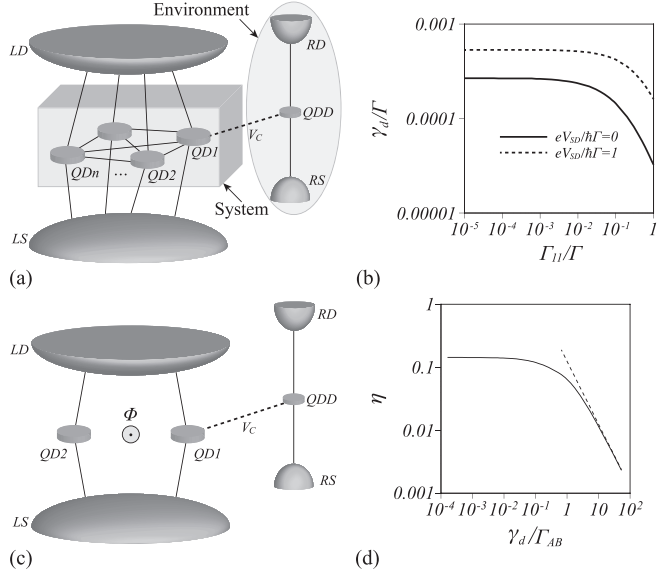}
\caption{\label{fig1} (a) Schematic diagram of a coherent system (indicated by the gray box), which consists of $n$ QDs coupled to source (LS) and drain (LD) reservoirs. QD1 couples to an environment (indicated by the gray ellipse). (b) Backaction dephasing rate $\gamma_d$ vs. linewidth function of QD1 $\Gamma_{11}$ for two values of bias applied to the QDD. (c) Schematic diagram of an AB interferometer containing two QDs with an environment containing a QDD. $\Phi$, and $V_C$ are the magnetic flux threading through an AB interferometer, and the Coulomb interaction between QD1 and the QDD, respectively. (d) Visibility of AB oscillations $\eta$ in the linear conductance through an AB interferometer as a function of the dephasing rate normalized by the linewidth function of QD1 and QD2 $\Gamma_{AB}$ when $\epsilon_1=\epsilon_2=0, V_C/\hbar\Gamma=0.05, \alpha=0.5$, and $T=0$.}
\end{figure}

Here we define the backaction dephasing induced by coupling with an environment. As shown in Fig. \ref{fig1}(a), we consider a \textit{coherent system} that includes $n$ QDs coupled to source (LS) and drain (LD) reservoirs. Moreover, we assume that only QD1 in the system couples to the environment as the result of Coulomb interaction whose strength is denoted by $V_C$. Then,  on the basis of the localized state in each QD, the retarded Green's function of the system is given by
\begin{eqnarray}
[G_{ij}^r(\epsilon)]^{-1}&=&\left\{\begin{array}{cc}
    \left[g_{ii}^r(\epsilon)\right]^{-1}-\delta_{i1}\Sigma_{ii}^r(\epsilon) & (i=j)  \\
    \left[g_{ij}^r(\epsilon)\right]^{-1} & (i\neq j)  \\ 
  \end{array}\right.,\label{green}
\end{eqnarray}
where $g_{ii}^r(\epsilon)$ is the retarded Green's function of the noninteracting $i$th QD tunnel coupled to reservoirs, and $g_{ij}^r(\epsilon)$ includes the direct and indirect inter-dot couplings between the $i$th and $j$th QD \cite{kubo}. The imaginary part of the retarded self-energy of the QD1, $\mbox{Im}\{\Sigma_{11}^{r}(\epsilon)\}=-\gamma_d(\epsilon)$, describes the backaction dephasing induced by Coulomb coupling with the environment. We employ the nonequilibrium second-order perturbation theory for $V_C$ \cite{2nd,interpolative}. In general, the real part of $\Sigma_{11}^r(\epsilon)$ provides the energy level shift of QD1. However, to discuss only the dephasing effect we compensate for this energy level shift \cite{kubo-dephasing}. In the following, we assume that the coherent system is always in the linear transport regime. Thus, we focus on the Fermi energy as the incident energy of an electron and consider $\gamma_d(\epsilon=0)$ in our calculation of backaction dephasing at low temperatures, where we choose the Fermi energy as the origin of the energy.

As an environment (see Fig. \ref{fig1}(a)), we consider a QDD that is tunnel coupled to source (RS) and drain (RD) reservoirs by coupling strengths (linewidth functions) $\Gamma_{RS}$ and $\Gamma_{RD}$ ($\Gamma_{RS}+\Gamma_{RD}=\Gamma$) to measure the current through a QDD, and that couples to QD1 capacitively. A source-drain bias voltage $V_{SD}$ is applied to the QDD. Here the $V_{SD}$ is defined as the electrochemical potential difference between the reservoirs RS and RD, namely $\mu_{RS}-\mu_{RD}=eV_{SD}$ where $\mu_{RS}=\lambda eV_{SD}$ and $\mu_{RD}=(\lambda-1)eV_{SD}$ with $0\le\lambda\le1$. The $\lambda$ is an experimentally tunable parameter, which provides how to apply the bias voltage. We introduce $\Gamma_{11}$ as the coupling strength between QD1 and the reservoirs, and plot the $\Gamma_{11}$-dependence of the backaction dephasing rate in Fig. \ref{fig1}(b) when $\epsilon_1=0$, $\epsilon_{QDD}=0$, $V_C/\hbar\Gamma=0.05$, and $k_BT/\hbar\Gamma=0.01$, where $\epsilon_1$, $\epsilon_{QDD}$, and $T$ are the energy level of QD1, the energy level of the QDD, and the temperature, respectively. When there is no direct inter-dot tunnel coupling and $\Gamma_{11}=0$, the $\gamma_d$ value corresponds to the backaction dephasing rate induced by an environment when QD1 is decoupled from the reservoirs \cite{which-path1}. As $\Gamma_{11}$ increases, the backaction dephasing rate decreases slowly in the small $\Gamma_{11}$ regime and rapidly in the large $\Gamma_{11}$ regime. The monotonic decrease in the backaction dephasing rate in the large $\Gamma_{11}$ regime can be explained as follows. The dwell time of an electron in QD1 is given by $\tau_{11}^{{\small \mbox{dwell}}}=(\Gamma_{11})^{-1}$. An increase in $\Gamma_{11}$ leads to a decrease in $\tau_{11}^{{\small \mbox{dwell}}}$. A similar result was obtained in Ref. \onlinecite{dephasing-qdd2} in relation to inter-dot tunnel coupling. As a result, the effective interaction time of the electrons passing through QD1 with the QDD becomes shorter, and the backaction dephasing rate decreases.

Although our results are general, we focus on an AB interferometer as a coherent system to observe backaction dephasing effects experimentally \cite{which-path6}. As shown in Fig. \ref{fig1}(c), we consider an AB interferometer containing two QDs (QD1 and QD2) and two Fermi liquid reservoirs LS and LD. Only QD1 couples to a QDD in an environment.  We assume that the level spacing is much larger than other energy scales, and consider only a single energy level in each QD. To focus on the coherent charge transport, we neglect the spin degree of freedom. In the tunneling amplitude between the reservoirs and QDs in an AB interferometer, we introduce the AB phase $\phi=2\pi\Phi/\Phi_0$ as a Peierls phase factor, where $\Phi$ is the magnetic flux threading through an AB interferometer and $\Phi_0=h/e$ is the magnetic flux quantum. We define the tunneling amplitude between the reservoir $\nu$ and the $j$th QD by $t_{\nu k}^{(j)}(\phi)$. From the definition of the linewidth function, $\Gamma_{ij}^{\nu}(\epsilon,\phi)=(2\pi/\hbar)\sum_k{t_{\nu k}^{(i)}}^*(\phi)t_{\nu k}^{(j)}(\phi)\delta(\epsilon-\epsilon_{\nu k})$, the diagonal functions are $\Gamma_{ii}^{\nu}(\epsilon)=(2\pi/\hbar)\sum_k|t_{\nu k}^{(i)}|^2\delta(\epsilon-\epsilon_{\nu k})$, where $\nu\in\{LS,LD \}$ and $i=1,2$. In our evaluation of the off-diagonal functions, we take account of the wave number dependence of the tunneling amplitude, and we obtained $\Gamma_{12}^{LD}(\phi)/\Gamma_{12}^{LS}(\phi)=(\alpha_{LD}/\alpha_{LS})\sqrt{\Gamma_{11}^{LD}\Gamma_{22}^{LD}/\Gamma_{11}^{LS}\Gamma_{22}^{LS}}e^{i\phi}$, $\Gamma_{21}^{LS}(\phi)=[\Gamma_{12}^{LS}(\phi)]^*$, and $\Gamma_{21}^{LD}(\phi)=[\Gamma_{12}^{LD}(\phi)]^*$. Here we introduce the coherent indirect coupling parameter $\alpha_{\nu}$, which characterizes the indirect coupling strength between two QDs via the reservoir $\nu$ \cite{kubo,hatano}. Moreover, we assume the wide-band limit, namely we neglect the energy dependence of the linewidth function: $\Gamma_{ij}^{\nu}(\epsilon,\phi)\equiv\Gamma_{ij}^{\nu}(\phi)$. Using these definitions, we calculate the linear conductance $G_{AB}$ through an AB interferometer with the Green's function technique \cite{current}. However, in the nonequilibrium second-order perturbation theory it is well-known that the current conservation is violated \cite{2nd}. Thus, we employ the interpolative approach, which satisfies the current conservation \cite{interpolative}. In a weak interaction regime, the imaginary part of the retarded self-energy described by the interpolative second-order perturbation theory behaves qualitatively the same as that described by the conventional second-order perturbation theory. To clarify the physical meaning of the backaction dephasing, we show the retarded self-energy described by the conventional second-order perturbation theory to analyze of the backaction dephasing. We calculate the conducetance with retarded self-energy by the interpolative second-order perturbation theory since we have to consider current conservation. We define the visibility of AB oscillations as $\eta\equiv(G_{AB}^{{\small \mbox{max}}}-G_{AB}^{{\small \mbox{min}}})/(G_{AB}^{{\small \mbox{max}}}+G_{AB}^{{\small \mbox{min}}})$ using the maximum and minimum values of the linear conductance $G_{AB}$. In the following, we consider the symmetric situation where $\Gamma_{11}^{LS(LD)}=\Gamma_{22}^{LS(LD)}\equiv\Gamma_{AB}/2$ and $\alpha_{\nu}\equiv\alpha$. The backaction dephasing rate dependence of the visibility of AB oscillations is shown in Fig. \ref{fig1}(d) when $\epsilon_{AB}=0$, $V_C/\hbar\Gamma=0.05$, $\alpha=0.5$, $T=0$, where the energy levels of QD1 and QD2 are defined as $\epsilon_1=\epsilon_2\equiv\epsilon_{AB}$. As the backaction dephasing rate increases for fixed $\Gamma_{AB}$, the visibility of the AB oscillations decreases since the interference is suppressed by the decoherence effect from the environment. Similarly, for a fixed $\gamma_d$, when $\tau_{11}^{{\small \mbox{dwell}}}=1/\Gamma_{AB}$ increases, the visibility of the AB oscillations decreases monotonically. At the limit of $\gamma_d\to 0$, the visibility approaches the value where there is no coupling to an environment. In contrast, the backaction dephasing rate is expressed by $\gamma_d\simeq\gamma_d^{(0)}+\gamma_d^{(1)}\Gamma_{AB}$ as a polynomial function of $\Gamma_{AB}$ in the weak QD-reservoir coupling regime. Under this condition, the visibility of the AB oscillations is given by $\eta\simeq(\alpha^2/2)[\gamma_d^{(0)}/\Gamma_{AB}]^{-1}$ which is shown by the dotted line in Fig. \ref{fig1}(d). 

We consider the physical origin of the backaction dephasing induced by the QDD. The backaction dephasing rate is expressed as
\begin{eqnarray}
\gamma_d(\epsilon=0)&=&\frac{1}{2}\left(\frac{V_C}{\hbar} \right)^2\int\frac{dE_1}{\hbar}[\rho_{11}(E_1)+\rho_{11}(-E_1)]\nonumber\\
&&\times[1-f_{eq}(E_1)]S_{nn}(-E_1).\label{charge_backaction}
\end{eqnarray}
Here $\rho_{11}(\epsilon)$ is the density of states of QD1 when there is no coupling between QD1 and the QDD, $f_{eq}(\epsilon)=1/[e^{\epsilon/k_BT}+1]$ is the equilibrium Fermi-Dirac distribution function, and $S_{nn}(\hbar\omega)\equiv\int dt\langle \delta n_{QDD}(t)\delta n_{QDD}(0) \rangle e^{i\omega t}$ is the spectral \textit{charge noise} in a QDD. Here $\delta n_{QDD}(t)\equiv n_{QDD}(t)-\langle n_{QDD}(t) \rangle$, where $n_{QDD}(t)$ is the number operator of a QDD. From Eq. (\ref{charge_backaction}), the $\gamma_d$ is symmetric with repsect to the QD energy $\epsilon_{AB}$ in an AB interferometer. Eq. (\ref{charge_backaction}) includes only the unsymmetrized spectral charge noise which denotes the unidirectional energy transfer from an environment to the system in the inelastic scattering process \cite{imry}. To clarify this point, in the linear transport regime, we rewrite the $\gamma_d$ as
\begin{eqnarray}
\gamma_{d,linear}(\epsilon=0)&=&\frac{1}{2}\left(\frac{V_C}{\hbar} \right)^2\left(\frac{\Gamma_{RS}+\Gamma_{RD}}{\Gamma_{RS}\Gamma_{RD}} \right)^2\nonumber\\
&&\times\int\frac{dE_1}{\hbar}[\rho_{11}(E_1)+\rho_{11}(-E_1)]\nonumber\\
&&\times[1-f_{eq}(E_1) ]\nonumber\\
&&\times\int\frac{dE_2}{2\pi\hbar}f_{eq}(E_1+E_2)[1-f_{eq}(E_2) ]\nonumber\\
&&\times\mathcal{T}_{QDD}(E_1+E_2)\mathcal{T}_{QDD}(E_2),\label{backaction_1}
\end{eqnarray}
where $\mathcal{T}_{QDD}(\epsilon)$ is the transmission probability through a QDD. Equation (\ref{backaction_1}) describes the inelastic electron-electron scattering process between QD1 and a QDD. The three factors including the Fermi-Dirac distribution function $1-f_{eq}(E_1)$, $f_{eq}(E_1+E_2)$, and $1-f_{eq}(E_2)$ correspond to the probabilities of the final state in an AB interferometer, and the initial and final states in an environment in the inelastic scattering process, respectively. Such dephasing corresponds to a relaxation caused by the inelastic electron-electron scattering process in Fermi liquid theory \cite{fermi-liquid}.

\begin{figure}
\includegraphics[scale=0.43]{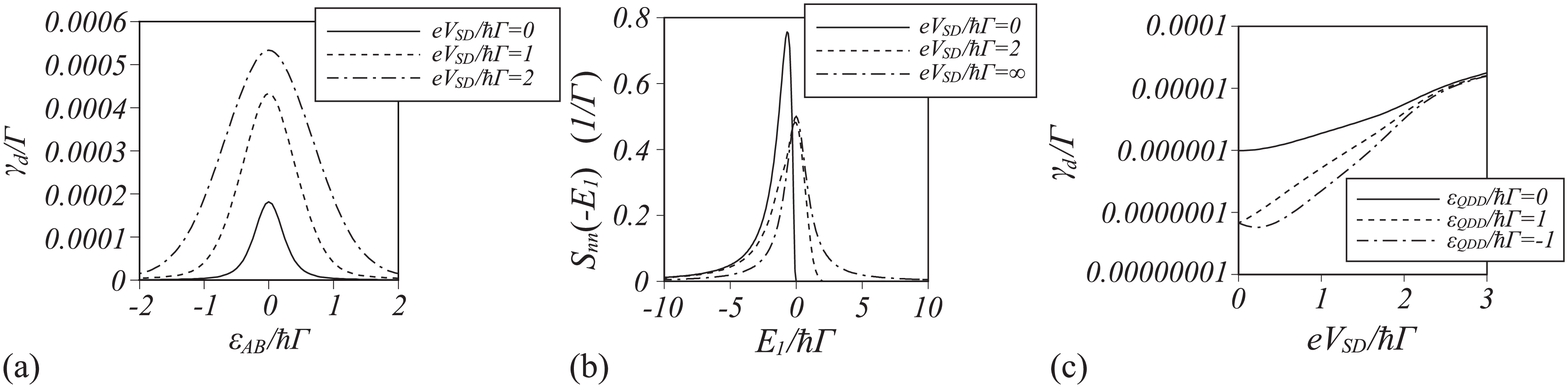}
\caption{\label{fig2} Backaction dephasing rate and spectral charge noise of a QDD when $V_C/\hbar\Gamma=0.05$, $\Gamma_{AB}/\Gamma=0.1$, $\alpha=0.5$, and $\phi=0$. (a) Backaction dephasing rate $\gamma_d$ vs. QD energy $\epsilon_{AB}$ for various bias voltages when $\Gamma_{RS}=\Gamma_{RD}=\Gamma/2$. The solid, dashed, and dash-dotted lines indicate the condition of $V_{SD}=0$, $eV_{SD}/\hbar\Gamma=1$, and $eV_{SD}/\hbar\Gamma=2$, respectively. (b) Spectral charge noise of a QDD at $\epsilon_{QDD}=0$, $T=0$, and $\Gamma_{RS}=\Gamma_{RD}=\Gamma/2$ for various bias voltages. The solid, dashed, and dashed-dotted lines indicate the condition of $V_{SD}=0$, $eV_{SD}/\hbar\Gamma=2$, and $eV_{SD}/\hbar\Gamma=\infty$. (c) Bias voltage dependences of the backaction dephasing rate at zero temperature for various QDD energy levels for $\epsilon_{AB}/\hbar\Gamma=2$, $\Gamma_{RS}/\Gamma=0.9$, $\Gamma_{RD}/\Gamma=0.1$, and $k_BT/\hbar\Gamma=0.1$. The solid, dashed, and dash-dotted lines indicate the condition of $\epsilon_{QDD}=0$, $\epsilon_{QDD}/\hbar\Gamma=1$, and $\epsilon_{QDD}/\hbar\Gamma=-1$, respectively.}
\end{figure}

Figure \ref{fig2}(a) depicts the $\epsilon_{AB}$-dependence of the backaction dephasing rates for various bias voltages at $k_BT/\hbar\Gamma=0.1$ when $\epsilon_{QDD}=0$, $V_C/\hbar\Gamma=0.05$, $\Gamma_{AB}/\Gamma=0.1$, $\alpha=0.5$, $\phi=0$, $\Gamma_{RS}=\Gamma_{RD}=\Gamma/2$, and $\lambda=0.5$. The $\gamma_d$ is symmetric with respect to QD energy $\epsilon_{AB}$, and is intensified as the bias voltage $V_{SD}$ increases. To understand the increase of backaction by bias voltage, we consider the spectral charge noise at $T=0$. The spectral charge noise is non-zero for $E_1<eV_{SD}$ as shown in Fig. \ref{fig2}(b). Because of the factor $1-f_{eq}(E_1)$ in Eq. (\ref{charge_backaction}), the frequency regime of $E_1>0$ is relevant for backaction dephasing, and the charge noise increases in the relevant frequency regime as the bias voltage becomes larger. As a result, the backaction dephasing rate is strengthened as the bias voltage increases. In the following, we discuss that such monotonic increase is not universal behavior.

In general, for finite temperatures \cite{note}, the $\gamma_d$ can be expressed as polynomial functions of a bias voltage $V_{SD}$ as follows: $\gamma_d=\gamma_d^{(0)}+\gamma_d^{(1)}eV_{SD}/\hbar\Gamma+\gamma_d^{(2)}(eV_{SD}/\hbar\Gamma)^2+O(eV_{SD}/\hbar\Gamma)^3$, where $\gamma_d^{(0)}\neq 0$, $\gamma_d^{(1)}\neq 0$ except for $\lambda\Gamma_{RS}=(1-\lambda)\Gamma_{RD}$, and $\gamma_d^{(2)}$ is positive. Thus, the dephasing rate is given by the equilibrium charge noise at $V_{SD}=0$ for finite temperature $T$. If $\lambda\Gamma_{RS}\neq(1-\lambda)\Gamma_{RD}$ and $T>0$, the dephasing rate has a linear contribution in $V_{SD}$ in a low bias voltage regime ($eV_{SD}\ll\hbar\Gamma$). The increase or decrease of the backaction dephasing rate in the low bias voltage regime is determined by the sign of $\gamma_d^{(1)}$. Here we focus on the condition of $\Gamma_{AB}\ll\Gamma,k_BT/\hbar$, namely the QD1 is almost isolated and the $\rho_{11}(\epsilon,\phi)\simeq\hbar\delta(\epsilon-\epsilon_{AB})$. Then the sign of $\gamma_d^{(1)}$ is given by the sign of $[\lambda\Gamma_{RS}-(1-\lambda)\Gamma_{RD}]\epsilon_{QDD}$ and is independent of $\epsilon_{AB}$. For $T=0$ and $\lambda\Gamma_{RS}=(1-\lambda)\Gamma_{RD}$, the dephasing rate increases parabolically for $V_{SD}$. In Fig. \ref{fig2}(c), we plot the $V_{SD}$-dependence of the backaction dephasing rate at $k_BT/\hbar\Gamma=0.1$ for various QDD energy levels when $\epsilon_{AB}/\hbar\Gamma=2$, $V_C/\hbar\Gamma=0.05$, $\Gamma_{AB}/\Gamma=0.1$, $\alpha=0.5$, $\phi=0$, $\Gamma_{RS}/\Gamma=0.9$, $\Gamma_{RD}/\Gamma=0.1$, $k_BT/\hbar\Gamma=0.1$, and $\lambda=0.5$. Under this condition, the sign of $\gamma_d^{(1)}$ follows the sign of $\epsilon_{QDD}$ as shown in the low bias voltage regime. The backaction dephasing rate is saturated for the high bias voltage as shown in Fig. \ref{fig2}(c) since the spectral charge noise is saturated to $S_{nn}(-E_1)|_{V_{SD}\to\infty}=\frac{2\Gamma_{RS}\Gamma_{RD}}{\Gamma}\frac{1}{\left(\frac{E_1}{\hbar} \right)^2+\Gamma^2}$ at the high bias limit as shown in Fig. \ref{fig2}(b). To observe such behaviors of the backaction dephasing rate for the bias voltage $V_{SD}$ and QDD energy level $\epsilon_{QDD}$ experimentally, we have only to investigate the visibility of the AB oscillations in the linear conductance through an AB interferometer with referring the results in Fig. \ref{fig1}(d).

To summarize, we have shown that the Coulombic backaction of a QDD is due to charge noise unlike a QPC. The backaction dephasing rate is symmetric with respect to the QD energy level $\epsilon_{AB}$ in an AB interferometer. In the linear transport regime through a QDD, charge noise induced backaction dephasing is described by an inelastic electron-electron scattering process. Out-of-equilibrium, in the low bias voltage regime, the increase or decrease of dephasing rate depends on the QDD energy level, the linewidth functions, and how to apply the bias voltage, and the backaction dephasing rate is saturated for the high bias voltage regime. Such behaviors can be extracted experimentally from the bias voltage dependence of the visibility of the AB oscillations in the linear conductance.

We thank A. Aharony, O. Entin-Wohlman, S. Tarucha, and Y. Utsumi for useful discussions. Part of this work is supported financially by JSPS MEXT Grant-in-Aid for Scientific Research on Innovative Areas (21102003) and Funding Program for World-Leading Innovative R\&D Science and Technology (FIRST).


\begin{thebibliography}{99}
\bibitem{heisenberg} W. Heisenberg, \textit{The Physical Principles of the Quantum Theory} (University of Chicago Press, 1930).
\bibitem{qc} M. A. Nielsen and I. L. Chuang, \textit{Quantum Computation and Quantum Information} (Cambridge University Press, 2000).
\bibitem{loss} D. Loss and D. P. DiVincenzo, Phys. Rev. A \textbf{57}, 120 (1998).
\bibitem{which-path1} Y. Levinson, Europhys. Lett. \textbf{39}, 299 (1997).
\bibitem{which-path2} I. L. Aleiner, N. S. Wingreen, and Y. Meir, Phys. Rev. Lett. \textbf{79}, 3740 (1997).
\bibitem{which-path3} A. Silva and S. Levit, Phys. Rev. B\textbf{63}, 201309(R) (2001).
\bibitem{which-path4} K. Kang, Phys. Rev. Lett. \textbf{95}, 206808 (2005).
\bibitem{which-path5} C. E. Young and A. A. Clerk, Phys. Rev. Lett. \textbf{104}, 186803 (2010).
\bibitem{which-path6} E. Buks, R. Schuster, M. Heiblum, D. Mahalu, and V. Umansky, Nature \textbf{391}, 871 (1998).
\bibitem{which-path7} D. Sprinzak, E. Buks, M. Heiblum, and H. Shtrikman, Phys. Rev. Lett. \textbf{84}, 5820 (2000).
\bibitem{which-path8} D.-I. Chang, G. L. Khym, K. Kang, Y. Chung, H.-J. Lee, M. Seo, M. Heiblum, D. Mahalu, and V. Umansky, Nature Phys. \textbf{4}, 205 (2008).
\bibitem{qdd1} R. J. Schoelkopf, P. Wahlgren, A. A. Kozhevnikov, P. Delsing, and D. E. Prober, Science \textbf{280}, 1238 (1998).
\bibitem{qdd2} Y. Makhlin, G. Sch\"{o}n, and A. Shnirman, Rev. Mod. Phys. \textbf{73}, 357 (2001).
\bibitem{qdd3} W. Lu, Z. Ji, L. Pfelffer, K. W. West, and A. J. Rimberg, Nature \textbf{73}, 357 (2003).
\bibitem{qdd4} T. Fujisawa, T. Hayashi, Y. Hirayama, H. D. Cheong, and Y. H. Jeong, Appl. Phys. Lett. \textbf{84}, 2343 (2004).
\bibitem{qdd5} C. Barthel, M. Kj\ae rgaard, J. Medford, M. Stopa, C. M. Marcus, M. P. Hanson, and A. C. Gossard, Phys. Rev. B\textbf{81}, 161308(R) (2010).
\bibitem{set1} A. Aassime, G. Johansson, G. Wendin, R. J. Schoelkopf, and P. Delsing, Phys. Rev. Lett. \textbf{86}, 3376 (2001).
\bibitem{set2} G. Johansson, A. K\"{a}ck, and G. Wendin, Phys. Rev. Lett. \textbf{88}, 046802 (2002).
\bibitem{dephasing-qdd1} G. L. Khym, Y. Lee, and K. Kang, J. Phys. Soc. Jpn. \textbf{75}, 063707 (2006).
\bibitem{dephasing-qdd2} V. Moldoveanu, M. Tolea, and B. Tanatar, Phys. Rev. B \textbf{75}, 045309 (2007).
\bibitem{kubo-dephasing} T. Kubo, Y. Tokura, and S. Tarucha, J. Phys. A \textbf{43}, 354020 (2010).
\bibitem{kubo} T. Kubo, Y. Tokura, T. Hatano, and S. Tarucha, Phys. Rev. B \textbf{74}, 205310 (2006).
\bibitem{2nd} Selman Hershfield, John H. Davies, and John W. Wilkins, Phys. Rev. B \textbf{46}, 7046 (1992).
\bibitem{interpolative} A. Levy Yeyati, A. Mart\'{i}n-Rodero, and F. Flores, Phys. Rev. Lett. \textbf{71}, 2991 (1993).
\bibitem{hatano} T. Hatano, T. Kubo, Y. Tokura, S. Amaha, S. Teraoka, and S. Tarucha, Phys. Rev. Lett. \textbf{106}, 076801 (2011).
\bibitem{current} Y. Meir and N. S. Wingreen, Phys. Rev. Lett. \textbf{68}, 2512 (1992).
\bibitem{imry} Y. Imry, \textit{Introduction to Mesoscopic Physics} (Oxford University Press, 1997).
\bibitem{fermi-liquid} P. Nozi\`{e}res, \textit{Theory of Interacting Fermi Systems} (Benjamin, 1964).
\bibitem{note} At $T=0$, it is noted that the spectral charge noise is singular, and thus we cannot expand the $\gamma_d$ as a polynomial function of bias voltage.
\end{thebibliography}
\end{document}